\documentclass[aps,prb,twocolumn,superscriptaddress,showpacs]{revtex4}     
\usepackage{graphicx}
\usepackage{bm}
\usepackage{color}
\usepackage{amsmath}

\begin{document}

\title{Spintronic Mechanics of a Magnetic Nanoshuttle}

\author{Robert I. Shekhter}
\affiliation{Department of Physics, University of Gothenburg,
SE-412 96 G{\" o}teborg, Sweden}
\author{Artem Pulkin}
\affiliation{Department of Physics, University of Gothenburg,
SE-412 96 G{\" o}teborg, Sweden}
\author{Mats Jonson}
\email{mats.jonson@physics.gu.se}
\affiliation{Department of Physics, University of Gothenburg,
SE-412 96 G{\" o}teborg, Sweden} \affiliation{SUPA, Institute of Photonics and Quantum Sciences, 
Heriot-Watt University, Edinburgh EH14 4AS, Scotland, UK}
\affiliation{Department of Physics, Division of Quantum Phases and
Devices, Konkuk University, Seoul 143-701, Korea}

\date{\today}

\begin{abstract}
We investigate theoretically the prospects for using a magnetic nanoelectromechanical single-electron tunneling (NEM-SET) device as an electronic spin filter. We find that strong magnetic exchange forces on the net spin of the mobile central dot of the NEM-SET structure lead to spin-dependent mechanical displacements (``spin polarons"), which give rise to vastly different tunnelling probabilities for electrons of different spin. The resulting spin polarization of the current can be controlled by bias and gate voltages and be very close to 100 \% at voltages and temperatures below a characteristic correlation energy set by the sum of the polaronic and Coulomb blockade energies.
\end{abstract}


\maketitle

\noindent
The ability to generate spin-polarized currents is an essential prerequisite for spintronics applications \cite{spintronics}. Accordingly, many studies of possible spin-filters capable of producing a spin-polarized tunnel current between ferromagnetic or normal metal electrodes have been reported, involving e.g. thin layers of ferromagnetic \cite{Bex} or ferrimagnetic \cite{Bex2} insulators, semiconductor quantum dots \cite{Aex1}, C$_{60}$ molecules \cite{Aex2}, and carbon nanotubes \cite{Aex3}. Quite large spin polarization factors have been achieved at low temperatures, 44\% as in Ref.~\onlinecite{Bex} and 33\% as in Ref.~\onlinecite{Bex2} being typical values at 10 K. However, the degree of polarization is much lower at room temperature and the pursuit of better spin filters continues.

In this context it is interesting to consider nanoelectromechanical spin filters.
The discrete nature of the electronic charge leads to a strong coupling of mechanical and electronic degrees of freedom in nanoscale single-electron tunneling (SET) structures with movable parts, such as a SET transistor with a flexible central island.
The resulting nanoelectromechanics is rich in mesoscopic phenomena, affecting both the electronic and the mechanical subsystems (see Ref.~\onlinecite{our-review} for a review). 

A conventional approach to the implementation of spin controlled nanoelectromechanics is based on spin dependent tunneling between magnetic conductors. The probability for electrons to tunnel is different in the two spin channels simply because the electron density of states in a magnetic metal is spin dependent. Hence the tunnel current may be spin-polarized with a number of interesting consequences for the electromechanics of, e.g., magnetic ``shuttle" structures \cite{6-radic}.

Qualitatively different phenomena may occur in nanometer-sized tunnel structures, where short-range magnetic exchange forces can be comparable in strength to the long-range electrostatic forces between charged elements of the device. There is ample evidence that the exchange field can be several tesla a few nm from the surface of a ferromagnet \cite{Aex1,Aex2,Aex3,adams} and the exponential decay of the field means that the corresponding force on a single electron spin can be very large. These spin-dependent exchange forces can give rise to various ``spintro-mechanical" phenomena and will be of central importance here, where we will show that it may result in a nanoelectromechanical spin filter with a theoretical efficiency close to 100\%.

\begin{figure}[h]
\includegraphics[width=0.5\textwidth,height=\textheight,keepaspectratio]{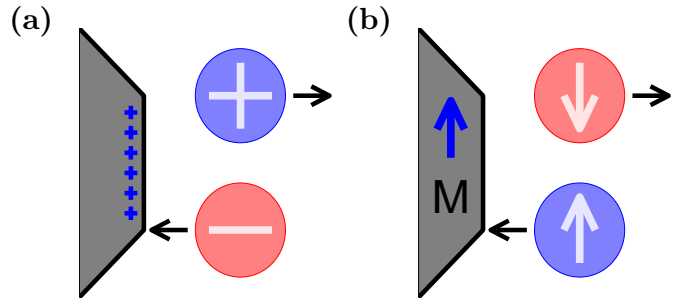}
\caption{\label{fig:spm} 
A movable quantum dot (circles) in a magnetic shuttle device can be displaced in response to two types of forces: (a) a long-range electrostatic force causing an electromechanical response if the dot has a net charge and (b) a short-range magnetic exchange force leading to a ``spintromechanical" response if the dot has a net magnetization (spin). The direction of the forces and displacements (arrows) depends on the relative signs of the charge and the magnetization, respectively. }
\end{figure}


The theory will be developed for a generic nanoelectromechanical ``shuttle" device in the form of a SET transistor with a central island that is movable along the line between a source- and a drain electrode as indicated in Fig.~\ref{fig:spm} (where only one electrode is shown). However, the theory applies equally well to, e.g., an extended island in he form of a suspended carbon nanotube in tunneling contact with two fixed electrodes and probed by an STM tip and this is the set-up we will use for our quantitative analysis. 
We assume that the island is a ``quantum dot" in the sense that spatial quantization only allows a single electron level to be populated at small bias voltages \cite{comment}. In this case both the electronic energy level on the island and the probability for electrons to tunnel to and from the bulk electrodes are affected by a mechanical displacement of the island. An electron that tunnels onto the dot changes both the charge and (by its spin) the total magnetization of the dot. The dot charge couples to the electric field associated with the bias voltage and leads to an electrostatic force (see Fig.~\ref{fig:spm}a) that acts on the movable dot with an electromechanical ``shuttle instability" \cite{2-gorelik} as a possible consequence. In a magnetic shuttle device the dot magnetization (spin) in addition couples to the magnetization of any nearby magnetic lead (or gate). This coupling gives rise to a short range exchange force (see Fig.~\ref{fig:spm}b), which causes a spintronically induced mechanical response of the dot and hence one may talk about the spintro-mechanics of a magnetic shuttle device.

In this Letter we will focus on one particular spintromechanical effect, viz. the formation of what we shall call ``spin-polaronic states" in a magnetic shuttle device.
%
%
In the limit of strong spintro-mechanical coupling the result of a tunneling event that changes the net spin on the dot is a polaronic modification of the mechanical vibrational states of the dot. The effect is similar to the modification caused by the addition of an electric charge on the dot \cite{7-glazman,8-wingreen,krive,10-braig}. In the latter case there is a Franck-Condon shift of the dot's position \cite{10-braig}, while in the former case the dot is displaced by the exchange force that appears as a spin is added in a combination of events that can be viewed as the formation of a spin-polaronic state.
Since the tunneling matrix element is exponentially sensitive to the position of the dot one expects a spin-dependent exponential renormalization of the tunneling probability caused by the formation of these spin-polaronic states.
This is the origin of the exponentially strong spin-dependent tunneling effect which we will be discussed in detail below.

\begin{figure}[h]
\includegraphics[width=0.3\textwidth,height=\textheight,keepaspectratio]{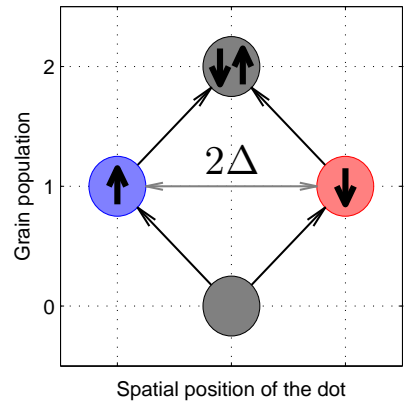}
\caption{\label{fig:4}
Diagram showing how the equilibrium position of the movable dot in Fig.~\ref{fig:spm}b depends on its net charge and spin.
The difference in spatial displacement discriminates transport through a single-occupied dot with respect to spin.}
\end{figure}

The described spintro-mechanical effect on the shuttle device can be tuned by changing the bias voltage in order to inject $n$ extra electrons onto the dot, thereby changing its spin as well as its charge. The diagram in Fig.~\ref{fig:4} shows how the resulting shift of the equilibrium position of the movable dot (due to the exchange force) depends on its spin state. 
%
If the bias voltage is increased in order to go from the $n=0$ to the $n=1$ state by adding one electron to the dot, the dot position will shift in different directions with respect to the leads depending on whether the spin of this electron is up or down. This spin-dependent shift lifts the spin-degeneracy of the probability for tunneling, which will be exponentially different for electrons with opposite spins. 
To see the effect of this spin discrimination of tunneling on the current, consider a magnetic shuttle device where the movable dot is closer to one of the leads (as indicated in Fig.~\ref{fig:spm}b), which therefore is more strongly exchange-coupled to the dot than the other lead. 
If an electron with spin anti-parallel to the lead magnetization tunnels onto the dot the exchange-force will repel the dot from the nearby lead and decrease its distance from the more distant lead, which in turn will enhance the probability for this electron to tunnel to that lead and contribute to the current. If an electron with parallel spin tunnels, the effect is the opposite and the probability for onward tunneling is decreased. The resulting current will therefore be spin polarized.

If the bias voltage is further increased to add another electron to the dot, the resulting $n=2$ state must be a spin singlet since we assume a single-level dot.  This spin-zero $n=2$ state would obviously not experience any exchange force. However, in the Coulomb blockade regime the $n=2$ state is energetically unfavourable, which is why (for an appropriate bias voltage) it can be expected to be frozen out even at relatively high temperatures. Spin-selective tunneling will therefore survive under the condition that Coulomb blockade of spintro-mechanical tunneling is preserved. 

So far we have argued that the injection of an extra electron onto the dot changes the exchange force \cite{footnote} on the dot with the result that its equilibrium position is shifted by a distance $\Delta$ (see Fig.~\ref{fig:4}).  We will refer to the extra electron plus the shift of the dot position as a spin polaron.
The Hamiltonian that describes the magnetic nanomechanical SET device has the standard form, except for its spin-dependent part (representing the magnetic exchange energy) which now depends on the mechanical displacement of the dot. Hence
\begin{equation}
H = H_{\rm leads} + H_{\rm tunnel} + H_{\rm dot}\,,
\label{eq:hamiltonian}
\end{equation}
where
\begin{equation}
H_{\rm leads} = \sum\limits_{\mathbf{k}, \sigma, s} a_{\mathbf{k} s \sigma}^\dagger a_{\mathbf{k} s \sigma} \epsilon_{\mathbf{k}s \sigma}
\end{equation}
describes electrons  (labeled by wave vector $\mathbf{k}$ and spin $\sigma=\uparrow, \downarrow$) in the two leads ($s=L$, $R$). 
Electron tunneling between the leads and the dot is modeled as  
\begin{equation}
H_{\rm tunnel} = \sum\limits_{\mathbf{k},\sigma, s}T_s \left ( x \right ) a_{\mathbf{k} s \sigma}^\dagger c_{\sigma} + H.c.,
\label{eq:htunnel}
\end{equation}
where the matrix elements $T_{s} (x)= T_{s}^{(0)} \exp(\mp x/\lambda)$, with $\lambda$ the characteristic tunneling length,
depend on the dot position $x$.

The movable single-level dot is modeled as a harmonic oscillator of angular frequency $\omega_0$,
\begin{equation}
\begin{array}{c}
H_{\rm dot} = \hbar \omega_0 b^\dagger b + \sum\limits_{\sigma} n_\sigma \left[ \epsilon_0 - {\rm sign}(\sigma) J \left ( x \right ) \right] + E_C n_{\uparrow} n_{\downarrow}\,,
\end{array}
\label{eq:hgrain}
\end{equation}
where ${\rm sign}(\uparrow,\downarrow)=\pm 1$, $E_C$ is the Coulomb energy associated with double occupancy of the dot and the eigenvalues of the electron number operators $n_\sigma$ is 0 or 1. The position dependent magnitude $J(x)$ of the spin dependent shift of the electronic energy level on the dot is due to the exchange interaction with the magnetic leads (and any external magnetic field). 

The strength of the polaronic coupling $\alpha = \Delta / x_0$ is determined by the ratio between the polaronic shift and the amplitude $x_0$ of the dot's quantum-mechanical zero-point oscillations. Another important parameter, $\beta = \Delta / \lambda$, is a measure of the effect of the 
spin-polaron formation on the probability for electron tunneling.
As we will see later large values of both these parameters lead to a highly spin polarized tunneling current. 

If the polaronic shift $\Delta$ is small compared to the characteristic length $l$ 
over which the exchange interaction changes significantly we may expand $J(x)$ to linear order in $x$ so that
\begin{equation}
J \left ( x \right ) = J^{\left ( 0 \right )} + j \, x
\label{eq:j}
\end{equation}
and without loss of generality furthermore assume that $J^{(0)} = 0$. In this case $\Delta =\vert j\vert  x_0^2/(\hbar\omega_0)$ and hence 
 \begin{equation}
\alpha = \frac{\vert j\vert x_0}{\hbar \omega_0} \quad {\rm and} \quad \beta = \alpha \frac{x_0}{\lambda}.
\label{eq:alpha}
\end{equation}

A full solution of the problem at hand can be obtained by solving the Liouville equation for the density matrix for both the electronic and vibronic subsystems.
Two different limits determine different scenarios for the nanoelectromechanical response of the device.
In the limit of low mechanical dissipation (high $Q$-factor) energy supplied by the external battery (used to maintain the bias voltage) may accumulate in the mechanical subsystem and eventually lead to an electromechanical instability and the onset of shuttle vibrations \cite{2-gorelik}.
In the opposite limit of strong mechanical dissipation (low $Q$-factor), which we will consider here, the vibronic subsystem is kept in equilibrium 
at the ambient temperature $T$.

In the weak tunneling limit a kinetic description of the electronic subsystem in terms of the probabilities $P_{\gamma \delta}$ to occupy the on-dot electronic states is possible ($\gamma$ and $\delta$ are the eigenvalues of $n_{\uparrow}$ and $n_{\downarrow}$, respectively).
By assuming that $P_{\gamma \delta}$ only changes through single-electron tunneling events between the dot and one or the other of the two leads, one arrives at the rate equation
\begin{equation}
\label{eq:kinetic}
\begin{split}
\frac{\partial}{\partial t}P_{\gamma \delta} =& \sum\limits_{s}\sum\limits_{\delta' = 1, 0} \left \lbrace \Gamma_{\gamma \delta; \gamma \delta'}^s P_{\gamma \delta'} - \Gamma_{\gamma \delta'; \gamma \delta}^s P_{\gamma \delta} \right \rbrace +
\\
& \sum\limits_{s} \sum\limits_{\gamma' =  1, 0} \left \lbrace \Gamma_{\gamma \delta; \gamma' \delta}^s P_{\gamma' \delta} - \Gamma_{\gamma' \delta; \gamma \delta}^s P_{\gamma \delta} \right \rbrace .
\end{split}
\end{equation}
Here the tunneling rates $\Gamma$ are are proportional to the squared modulus of the tunnel matrix elements between different stationary vibrational and electronic states. We are interested in the DC current through the device, which is determined by the time-independent solution of the kinetic equation (\ref{eq:kinetic}).  The spin-up current, e.g., can be expressed as
\begin{equation}
\label{eq:current}
I_{\uparrow} = e \left ( \Gamma^L_{0 0; 1 0}P_{1 0} +  \Gamma^L_{0 1 ; 1 1}P_{1 1} - \Gamma^L_{1 0 ;0 0} P_{0 0} -  \Gamma^L_{1 1; 0 1}P_{0 1} \right ).
\end{equation}

Although a general solution of Eqs.~(\ref{eq:kinetic}) and (\ref{eq:current}) for the stationary spin-up current is possible, the resulting expression  is quite complicated and not very suitable for a qualitative analysis.
This is why we will here restrict ourselves to the experimentally most likely case of an asymmetric tunneling device with $\Gamma^L_{\gamma\delta;\gamma'\delta'} \ll \Gamma^R_{\gamma\delta;\gamma'\delta'}$.
In this limit, where the electronic systems on the dot and in the nearest (right) lead will be in thermal equilibrium, the solution for the occupation probabilities $P_{\gamma \delta}$ will be independent of both  $\Gamma^L$ and $\Gamma^R$, while the current will depend on the tunneling rate $\Gamma^L$ to the most remote (left) lead.
A direct calculation (following Ref.~\onlinecite{8-wingreen}) of the tunnel matrix element $\Gamma^L_{1 0; 0 0}$, for example, gives the result
\begin{equation}
\label{eq:gammaFinal}
\Gamma^L_{1 0,0 0} = \frac{W_L}{\hbar} \Sigma_{L ,\uparrow} \left ( \epsilon \right )
e^{\left [ N + \frac{1}{2} \right ] \left ( \frac{x_0^2}{\lambda^2} - \alpha^2 \right ) - \beta}.
\end{equation}
Here $\epsilon = \epsilon_0 - E_P / 2$, with $E_P=  \alpha^2 \hbar \omega_0$, is the polaronically modified energy level on the dot, 
$W_L = 2 \pi \nu_L \left | T_L^{\left ( 0 \right )} \right |^2$, with $\nu_L$ the electron states density at Fermi level in the left lead, is the tunneling-induced width of the dot level.
Furthermore, 
\begin{eqnarray}
\nonumber
\Sigma_{L \sigma} \left ( \epsilon \right ) &=& \sum\limits_{n = -\infty}^{+\infty} f_n^L \cdot \exp \left ( \frac{n \hbar \omega_0}{2 k_B T} \right ) \left | \frac{ 1 + \lambda j / \hbar \omega_0}{1 - \lambda j / \hbar \omega_0} \right | ^{n} 
\\ \nonumber
&\times& I_n \left ( \frac{x_0^2}{\lambda^2} \left [ 1 - \left \{ \frac{j \lambda}{\hbar \omega_0} \right \}^2 \right ] \sqrt{ N(N + 1) } \right ),
\end{eqnarray}
where $I_n(...)$ is a Bessel function, $N$ is the average number of excited dot-vibration quanta,
\begin{equation}
N = 1/\left [\exp (\hbar\omega_0/k_BT) - 1 \right ],
\label{eq:N}
\end{equation}
and
$$f^L_n = 1/\left [\exp ([\epsilon + {\rm sign(\sigma)} n \hbar \omega_0-\mu]/k_BT) + 1 \right ].$$ 

Equation (\ref{eq:current}) for the current with the tunneling rates $\Gamma^L_{\gamma\delta;\gamma'\delta'}$ given by expressions like (\ref{eq:gammaFinal}) and the occupation probabilities $P_{\delta\gamma}$ obtained from the kinetic equation (\ref{eq:kinetic}) gives an analytical expression for the current.
At low temperatures and low bias voltages the current is affected by both the familiar (spin-independent) polaronic blockade of tunneling (through the parameter $\alpha$) \cite{7-glazman,8-wingreen,krive,10-braig,11-koch} and by spin-selective tunneling (through $\beta$).
For example, at $V \approx 0$ and $T\approx 0$ the partial current of electrons with spin $\sigma$ is 
\begin{equation}
I_\sigma \sim e \frac{W_L}{\hbar} \exp \left ( \frac{1}{2} \left [ \frac{x_0^2}{\lambda^2} - \alpha^2 \right ] - {\rm sign}(\sigma)\beta \right ),
\end{equation}
where a spin-independent prefactor of order 1 has been omitted. 

An increase of voltage or temperature lifts the polaronic blockade of tunneling when ${\rm max}\{eV/2, k_BT\}>E_P$. Tunneling will still be spin selective, however, as long as ${\rm max}\{eV/2, k_BT\}<E_P+E_C$ so that double occupancy of the dot is prevented by the Coulomb blockade. In this case the expression for the spin currents take the form
\begin{equation}
I_\sigma \sim e \frac{W_L}{\hbar} \exp \left ( \left [ 2 N + 1 \right ] \frac{x_0^2}{\lambda^2} - 2 \,{\rm sign}(\sigma)\beta \right ),
\label{eq:Isigma2}
\end{equation}
where $N$ is given by Eq.~(\ref{eq:N}).
At even higher voltages or temperatures, ${\rm max}\{eV/2, k_BT\}>E_P+E_C$, the Coulomb blockade of double dot occupancy is also lifted. A doubly occupied dot level will have no net spin and therefore the exchange force on the dot vanishes, no spin-polaron is formed, the spin currents are independent of spin projection, $\beta=0$ in Eq.~(\ref{eq:Isigma2}), and the spin-polaronic stimulation of the current ends with the result that the current drops at $eV \sim 2(E_P+E_C)$ (at low temperatures).
For higher voltages the current will be $\propto\exp(2T/T^*)$, where $k_BT^*=\hbar\omega_0/x_0^2$ (assuming $k_BT>\hbar\omega_0$). This is because the amplitude of the random thermal vibrations, which effectively reduce the tunneling distance, increases with temperature. 


The higher current at intermediate bias voltages is an example of ``Coulomb promotion of tunneling", which here depends on the spin-dependent shift of the dot position associated with the formation of spin polarons. Coulomb promotion of tunneling may also appear due to the standard spin dependent tunneling phenomenon (mentioned in the introduction) as discussed in Ref.~\onlinecite{12-gorelik}.

\begin{figure}[h]
\includegraphics[width=0.5\textwidth,height=\textheight,keepaspectratio]{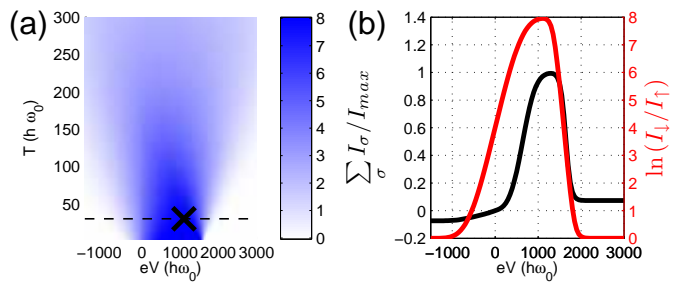}
\caption{\label{fig:plot}
Spin polarization of the current through our model magnetic NEM-SET device for $\hbar \omega_0/k_B = 0.01$~K, 
$\lambda = 10\, x_0 \sim 0.1$~nm, 
$\lambda j/k_B = 2$~K  [hence $E_P=400 \,\hbar\omega_0$] and $E_C=400 \,\hbar\omega_0$:
(a) $\ln(I_\downarrow / I_\uparrow$) as a function of bias voltage $V$ and temperature $T$ 
(b) $\ln(I_\downarrow / I_\uparrow$) as a function of $V$ at $T = 30\,\hbar \omega_0/k_B$ [grey (red) curve] and $I=I \left ( V \right )$ for the same parameters, i.e. along the dashed line in panel (a) [black curve].}
\end{figure}

For a numerical example we consider an extended quantum dot in the form of a single-wall carbon nanotube suspended between two electrodes (source) and probed by a magnetic-metal STM tip (drain), which is similar to the set-up used by LeRoy {\em et al.} \cite{LeRoy}. For a suspended tube length of order 1~$\mu$m the vibration frequency $\omega_0/2\pi$ of the fundamental bending mode is of order 100~MHz and hence $\hbar\omega_0\sim 1~\mu$eV ($\hbar \omega_0/k_B \sim 0.01$~K). Furthermore, assuming a diameter of 2 nm, the suspended mass $M$ of such a tube is such that the quantum oscillation amplitude $x_0\sim\sqrt{\hbar/2M\omega_0} \sim 0.01$~nm. The tunneling length is an atomic distance, so $\lambda\sim 0.1$~nm is a reasonable estimate and hence $\lambda \sim 10\,x_0 \gg x_0$. Finally, if one approximates the spatial gradient of the exchange field $j=\partial J(x)/\partial x \sim J/\lambda$ where $J\sim 0.1~$meV ($J/k_B\sim 1$~K) one finds that $\alpha \sim 10$ and $\beta \sim 1$. These parameter values are large enough for the predicted effect to be strong as is evident from Fig.~\ref{fig:plot}, where the current-voltage characteristics and spin polarization of the current are plotted as functions of voltage and temperature. In particular it is clear from Fig.~\ref{fig:plot} that the spin polarization of the current can be very close to $100\%$. The characteristic temperature $T^*$ is a few kelvin with these parameters.

One notes that the charging energy for the extended nanotube quantum dot of our example is rather small, 4~K, and hence well below room temperature. However, one may consider using functionalized nanotubes \cite{cnt.hybrids} or graphene ribbons \cite{graphene.hybrids} with one or more nm-sized metal or semiconductor nanocrystals attached. The small size of the crystals could give several orders of magnitude larger charging energies while not much affecting the low mechanical vibration frequencies of the carbon resonators.

In conclusion we have demonstrated that a spintro-mechanical coupling caused by spatially non-homogeneous exchange interactions in magnetic nanoelectromechanical single-electron tunneling structures may result in a very strong spin dependence of the probability for electron tunneling. As a result the theoretical value of the spin polarization of the electrical current can essentially be 100\%. 
For future work we note that the strong spin-dependence of mechanical displacements demonstrated here in the limit of strong mechanical dissipation, may lead to nontrivial spin-mechanical dynamics in the weak dissipation limit if both the spin dynamics and the mechanics is coherent.

{\em Acknowledgement.} Useful discussions with I. V. Krive and S. I. Kulinich as well as financial support from the European Commission (FP7-ICT-FET-225955 STELE), the Swedish VR, and the Korean WCU program
funded by MEST/NFR (R31-2008-000-10057-0) is gratefully acknowledged.


\begin{thebibliography}{99}
\bibitem{comment}
This assumption is made for convenience and is not essential since the Coulomb blockade ensures that only a single electron can enter the dot for low bias voltages and low temperatures.
\bibitem{footnote}
The effect of the much smaller change of the electrostatic force is neglected.
\end{thebibliography}
\end{document}